\documentclass[aps,secnumarabic,nofootinbib,amsmath,amssymb,superscriptaddress,nobibnotes,floatfix]{revtex4-1}
\usepackage{lipsum}
\usepackage{float}
\usepackage[squaren]{SIunits}
\usepackage{verbatim}
\usepackage{esint}
\usepackage{mathrsfs}
\usepackage{stmaryrd}
\usepackage{comment}
\usepackage{braket}
\usepackage{bbold}
\usepackage{xcolor}
\usepackage[latin1]{inputenc}
\usepackage{graphicx}
\usepackage{dsfont}
\usepackage{amsmath,amsfonts,amssymb,amsthm}
\usepackage{xspace}
\usepackage{times}
\usepackage{longtable}
\usepackage{extarrows}
\usepackage{ulem}   
\usepackage[colorlinks, linkcolor=blue, anchorcolor=blue, citecolor=blue]{hyperref}

\begin{document}


\title{Tracking quantum state evolution by the Berry curvature with a two-level system}

\author{Ze-Lin Zhang}
\affiliation{Fujian Key Laboratory of Quantum Information and Quantum Optics, College of Physics and Information Engineering, Fuzhou University, Fuzhou 350116, China}
\affiliation{School of Astronomy and Space Science, Nanjing University, Nanjing 210093, China}
\author{Ping Xu}
\affiliation{Fujian Key Laboratory of Quantum Information and Quantum Optics, College of Physics and Information Engineering, Fuzhou University, Fuzhou 350116, China}
\author{Zhen-Biao Yang}
\email{E-mail address: zbyang@fzu.edu.cn}
\affiliation{Fujian Key Laboratory of Quantum Information and Quantum Optics, College of Physics and Information Engineering, Fuzhou University, Fuzhou 350116, China}

\begin{abstract}
  We investigate two kinds of topological structures $($sphere and torus$)$ spanned by the controlled parameters of a driven two-level system's Hamiltonian, and consider the connection between the structures and the system's dynamics. We discuss the Berry curvature obtained through the dynamical response method, show the certain physical and observable manifolds including the gapped region probed by integrating the Berry curvature, and demonstrate the system's state evolution can be tracked and manipulated by extracting the Berry curvature.
\end{abstract}

\pacs{}

\maketitle

\section{Introduction}
\label{sec: Introduction}

  The concept of geometric phase, especially the Berry phase, plays a pivotal role in broadening our understanding of fundamental significance of geometry in physics~\cite{Berry-1984}. It stems from manipulating the geometrical structure formed of the system's parameter space, which drives the adiabatic cyclic evolution of nondegenerate quantum eigenstates. The topological structure of a certain closed manifold formed by the Hamiltonian parameter space can be captured by the distribution of the Berry curvature. In general, the geometric structure of the parameter space $($i.e., a spin-1/2 particle in a magnetic field$)$ and the space of states in Hilbert space $\mathbb{C}^2$ $($i.e., the Bloch sphere$)$ are sphere structures. It is an alternative to build a torus manifold formed in the parameter space of the Hamiltonian of a driven two-level system. We show how the manifold can be controlled by adjusting the drive. We extract the Berry curvature using the dynamically non-adiabatic response method~\cite{Gritsev-2012} $($different from the conventional Berry curvature~\cite{Hasan-2010}, as described in Appendix~\ref{sec: concise comparison}$)$ and attach it to the system's evolution characterized by certain quantum state fidelity, showing the tight connection between them.

  As the physical meaning of the singularity can be considered as the Abelian $($non-Abelian case is not considered here~\cite{Sugawa-2018}$)$ monopole in parameter space, the Berry curvature can be viewed as the magnitude of magnetic field emerging from it. Such a Berry curvature can be viewed as an intuitive geometric monitor between the Hamiltonian parameter space and the Hilbert space, and it can be directly measured~\cite{Flaschner-2016}. The tight link between the parametric manifold and the system's state evolution helps to explore the dynamics of the system from a geometric point of view. It makes sense to track the evolution of quantum states by redesigning the geometrical quantities$-$such as the observable Berry curvature.

  The configuration of this paper proceeds as follows. In Sec.~\ref{sec: Topology in the specific state manifold}, we introduce the concepts of the Berry curvature and the first Chern number $($$\mathds{C}_1$$,$ topological invariant$)$ associated with their physical meaning in quantum theory. We then introduce the dynamical response method for characterizing them. In Sec.~\ref{sec: From Sphere to Torus}, we compare two different manifolds, sphere and torus, which are parameterized by the system's Hamiltonian. Based on this point, we show the difference between the conventional Berry curvature and the dynamical Berry curvature $($obtained by dynamical method, Berry curvature should be used unless otherwise specified$)$ associated with their first Chern number in the torus case. It is worth mentioning that the first Chern number $($an artifact of approximating the curvature with the dynamical response method$)$ which we calculated in our work is different to the mathematically defined first Chern number. To highlight the physical and geometric meaning of the Berry curvature, we compare it with the Gaussian curvature $($only have geometric meaning$)$. A simple geometric transformation from sphere to torus is also shown. To test the robustness of the quantum state evolution, we show the dynamics of the system decaying to the reservoir. In Sec.~\ref{sec: The connection between the Berry curvature and the quantum state evolution}, we discuss the correlations between the Berry curvature and the quantum states during the deformation process of the manifolds. We also describe the experimental feasibility of this theoretical method. Summary and prospects are presented in Sec.~\ref{sec: Concluding remarks}.

\section{Topology in the specific state manifold}
\label{sec: Topology in the specific state manifold}

  We first introduce the Berry curvature in a fundamentally mathematical background. Let us consider a set of parameters $\textbf{\textit{R}} = (\textit{R}^{\mu}, \textit{R}^{\nu}, \cdots)\in\mathcal{M}$, where $\mathcal{M}$ denotes the Hamiltonian parameters base manifold and acts over the Hilbert space. Vector appears bold in the full text. The Berry curvature $\textit{B}_{\mu\nu}$ has a direct connection to quantum geometric tensor $\textit{Q}_{\mu\nu}$ introduced to depict the manifolds of adiabatically connected wave functions $|\psi(\textbf{\textit{R}})\rangle$ $($for simplicity, we use $|\psi\rangle$ instead of it$)$~\cite{Provost-1980}:
  \begin{eqnarray}\label{quantum geometric tensor}
  \textit{Q}_{\mu\nu}= \langle\partial_{\mu}\psi|\partial_{\nu}\psi\rangle-\langle\partial_{\mu}\psi|\psi\rangle\langle\psi|\partial_{\nu}
  \psi\rangle
  \end{eqnarray}
  with $\partial_{\mu(\nu)}\equiv{\partial}/{\partial\textit{R}^{\mu(\nu)}}$. The Berry curvature is given by the antisymmetric imaginary part of the quantum geometric tensor:
  \begin{eqnarray}\label{antisymmetric part}
  \textit{B}_{\mu\nu}=-2\textrm{Im}[\textit{Q}_{\mu\nu}]=\partial_{\mu}\textit{A}_{\nu}-\partial_{\nu}\textit{A}_{\mu},
  \end{eqnarray}
  where $\textit{A}_{\mu(\nu)}=i\langle\psi|\partial_{\mu(\nu)}|\psi\rangle$ is the Berry connection~\cite{Simon-1983}.

  In analogy to classical electrodynamics, the local gauge-dependent Berry connection $\textit{A}_{\mu(\nu)}$ could not be physically observed, while the gauge-invariant $\textit{B}_{\mu\nu}$ may be related to a physical observable that reflects the local geometry property of the eigenstates in the parameter space. For a general two-level system, there only have two eigenstates $|{e}\rangle$ and $|{g}\rangle$ of the Hamiltonian $\hat{\textit{H}}$, with the corresponding higher energy $\textit{E}_{e}$ and lower energy $\textit{E}_{g}$, in describing the Berry curvature
  \begin{eqnarray}\label{Berry curvature1}
  \textit{B}_{\mu\nu}^{N}=-\textrm{Im}\frac{\langle g|\partial_{\mu}\hat{\textit{H}}|e\rangle\langle e|\partial_{\nu}\hat{\textit{H}}| g\rangle-
  \langle g|\partial_{\nu}\hat{\textit{H}}|e\rangle\langle e|\partial_{\mu}\hat{\textit{H}}| g\rangle}{(\textit{E}_e-\textit{E}_g)^2}\cr
  \end{eqnarray}
  and
  \begin{eqnarray}\label{Berry curvature2}
  \textit{B}_{\mu\nu}^{S}=-\textrm{Im}\frac{\langle e|\partial_{\mu}\hat{\textit{H}}|g\rangle\langle g|\partial_{\nu}\hat{\textit{H}}| e\rangle-
  \langle e|\partial_{\nu}\hat{\textit{H}}|g\rangle\langle g|\partial_{\mu}\hat{\textit{H}}| e\rangle}{(\textit{E}_g-\textit{E}_e)^2},\cr
  \end{eqnarray}
  where the quantum states $|{e}\rangle=(1,0)^T$ and $|{g}\rangle=(0,1)^T$ represent the excited state and the ground state of the two-level system, respectively. Eqs.~$($\ref{Berry curvature1}$)$ and $($\ref{Berry curvature2}$)$ indicate that degeneracies are some singularities $($gapped regions$)$ that contribute nonzero terms to topological invariants. Such as, the first Chern number $\mathds{C}_1$, a topological invariant defined by
  \begin{eqnarray}\label{Chern number}
  \mathds{C}_1=\frac{1}{2\pi}\varoiint_{\textit{S}^2}\textbf{\textit{B}}\cdot d{\textbf{\textit{S}}}= \frac{1}{4\pi}\varoiint_{\textit{S}^2}\textit{B}_{\mu\nu}d\textit{R}^{\mu}\wedge d\textit{R}^{\nu},
  \end{eqnarray}
  where $\textbf{\textit{B}}$ is the Abelian field strength $($the vector form of the Berry curvature $\textit{B}_{\mu\nu}$$)$ over a closed two-dimensional manifold $\textrm{S}^2$, where ${\textbf{\textit{S}}}$ is a vector normal to the surface. When the manifold is not closed, the surface integral turns into the Berry phase: $\phi_{B}=\iint_{\textrm{S}^2}{\textbf{\textit{B}}}\cdot d{\textbf{\textit{S}}}$. In the circumstances, the first Chern number $\mathds{C}_1=\pm1$, which can be viewed as the Abelian monopoles, sources $\rho_N$ $($$\mathds{C}_1=+1$$)$ and sinks $\rho_S$ $($$\mathds{C}_1=-1$$)$, in the parameter space~\cite{Zhang1-2017,Zhang2-2017}.

  Ref.~\cite{Schroer-2014} states that the Berry curvature can be extracted from the linear response of the driven two-level system to non-adiabatic manipulations $($deviations from adiabaticity$)$ of its Hamiltonian $\hat{\textit{H}}$ $($$\mu=\theta$, $\nu=\phi$$)$ which leads to the generalized force $\textit{F}_{\phi}=-\langle \psi(t)|\partial_{\phi}\hat{\textit{H}}|\psi(t)\rangle$. In an Abelian system, the force along the $\phi$-direction reads
  \begin{eqnarray}\label{general force}
  \textit{F}_{\phi}= \textrm{const}+\theta_{t}\cdot\textit{B}_{\theta\phi}+\mathcal{O}({\theta_{t}}^2),
  \end{eqnarray}
  which results from a parameter $\theta$ ramping with velocity $\theta_t$, and $\textit{B}_{\theta\phi}$ is a component of the tensor $\textit{B}_{\mu\nu}$. To neglect the nonlinear term $\mathcal{O}({\theta_{t}}^2)$, the system's parameters should be ramped slowly enough. The focus of interest here is the second term $\theta_{t}\cdot\textit{B}_{\theta\phi}$, which is the desired first-order reaction force $($analogous to the Lorentz force$)$~\cite{Berry-1993}. It is worth mentioning that, different from the conventional Berry curvature $($full geometric meaning$)$~\cite{Hasan-2010}, the Berry curvature defined here is in physics
  more intuitive and experimentally more accessible~\cite{Roushan-2014}, as described in Appendix~\ref{sec: concise comparison}.

\section{From Sphere to Torus}
\label{sec: From Sphere to Torus}

\begin{figure}[h]
    \centering
    \includegraphics[width=5.4in]{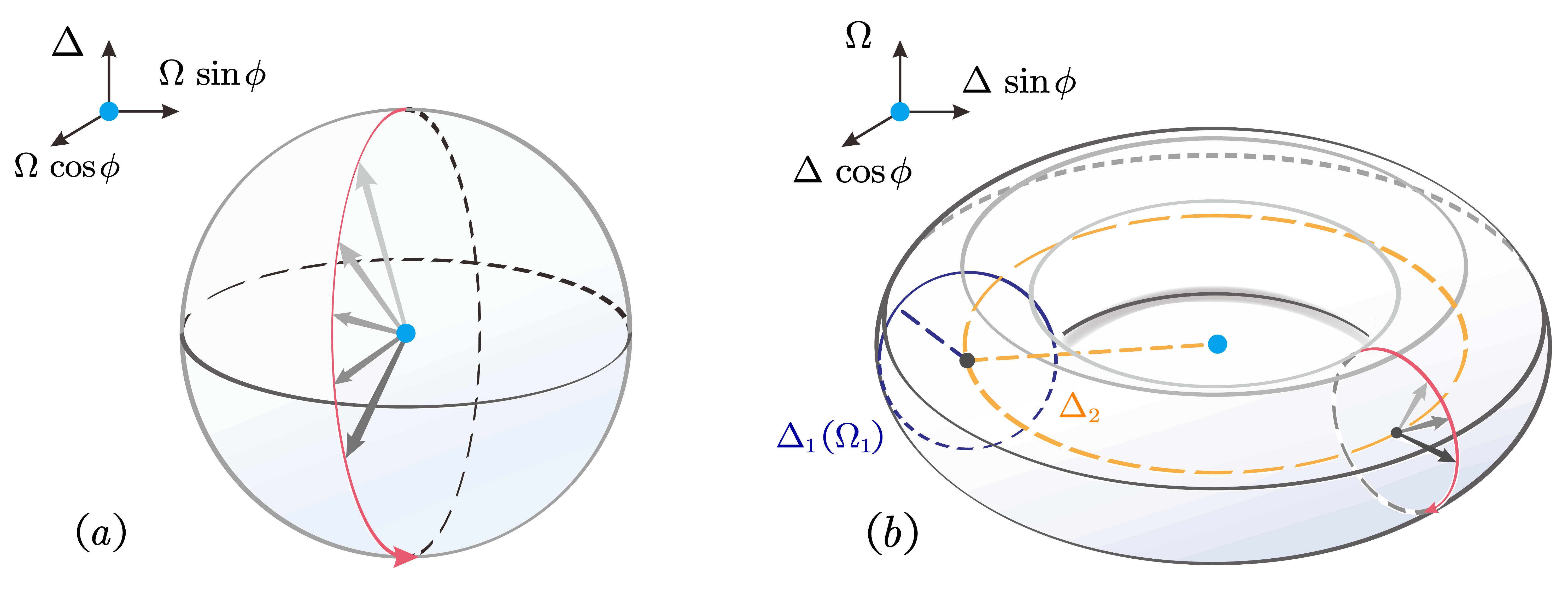}
    \caption{$($Color online$)$ $($a$)$ A schematic of a parameter sweep for $\Delta_2=0$, see Eq.~$(\ref{detuning})$. The sweep red line on the sphere depicts $\theta$ ramping from 0 to $\pi$. $($b$)$ The transformed case with $\Delta_2=2\Delta_1=2\Omega_1$. The sweep red line on the torus $(\textrm{T}^2)$ surface depicts $\theta$ ramping from 0 to $2\pi$. In this case, the radius of the blue circle $\Delta_1=\Omega_1$ is the half length of the orange circle's radius $\Delta_2$. The blue points shown in $($a$)$ and $($b$)$ depict the degenerate singularities $($gapped regions$)$.}
    \label{fig:figure1}
  \end{figure}

  Now, we assume a superconducting transmon qubit, which is effectively a nonlinear resonator $($with a transition frequency of $\omega_q$ and the reasonable anharmonicity, to ensure the transition only occurs between the first energy level $\textit{E}_g$ and the second energy level $\textit{E}_e$$)$, driven by a microwave drive. To investigate the Berry curvature characterized by the Hamiltonian parameter space and its connection to the evolution of the system's state, we perform a comparative study of two special kinds of manifolds, sphere and torus, in parameter space of the system's Hamiltonian, as shown in Fig.~\ref{fig:figure1}.

  We first consider the sphere case. In the rotating frame of the microwave drive with frequency $\omega_m$, and in the units with $\hbar=1$, the Hamiltonian for the qubit can be written as
  \begin{eqnarray}\label{Hamiltonian Sphere1}
  \hat{\textit{H}}_{\textrm{sph}}= \frac{1}{2}\left(
                   \begin{array}{cc}
                     \Delta & \Omega{\textit{e}}^{-\textit{i}\phi} \\
                     \Omega{\textit{e}}^{\textit{i}\phi} & -\Delta\\
                   \end{array}
                 \right),
  \end{eqnarray}
  where the detuning is $\Delta=\omega_m-\omega_q$, and $\phi$ is the phase of the drive. In principle, by changing the detuning $\Delta$ and the Rabi frequency $\Omega$, we could get arbitrary closed Hamiltonian manifold of the qubit. We set the detuning and the Rabi frequency to vary as
  \begin{eqnarray}\label{detuning}
  \Delta=\Delta_1\cos\theta+\Delta_2,~~~~
  \Omega=\Omega_1\sin\theta.
  \end{eqnarray}
  The corresponding parametric equations of Eq.~$($\ref{Hamiltonian Sphere1}$)$ is given by
  \begin{eqnarray}\label{para_sphere}
  \{\hat{\sigma}^x\}_{\textrm{sph}}=\Omega_1\sin\theta\cos\phi,~~~~
  \{\hat{\sigma}^y\}_{\textrm{sph}}=\Omega_1\sin\theta\sin\phi,~~~~
  \{\hat{\sigma}^z\}_{\textrm{sph}}=\Delta_1\cos\theta+\Delta_2,
  \end{eqnarray}
where $\{\cdot\}_{\textrm{sph}}$ represents the magnitude of vector component projection onto different axes of the three dimensional space of Hamiltonians generated by the Pauli matrices in the case of sphere, as shown in ~\ref{fig:figure1}$($a$)$. We notice that if we change the parameter $\Delta_2$, the whole sphere will move up and down in the parameter space.

  \begin{figure*}[t]
    \centering
    \includegraphics[width=6.8in]{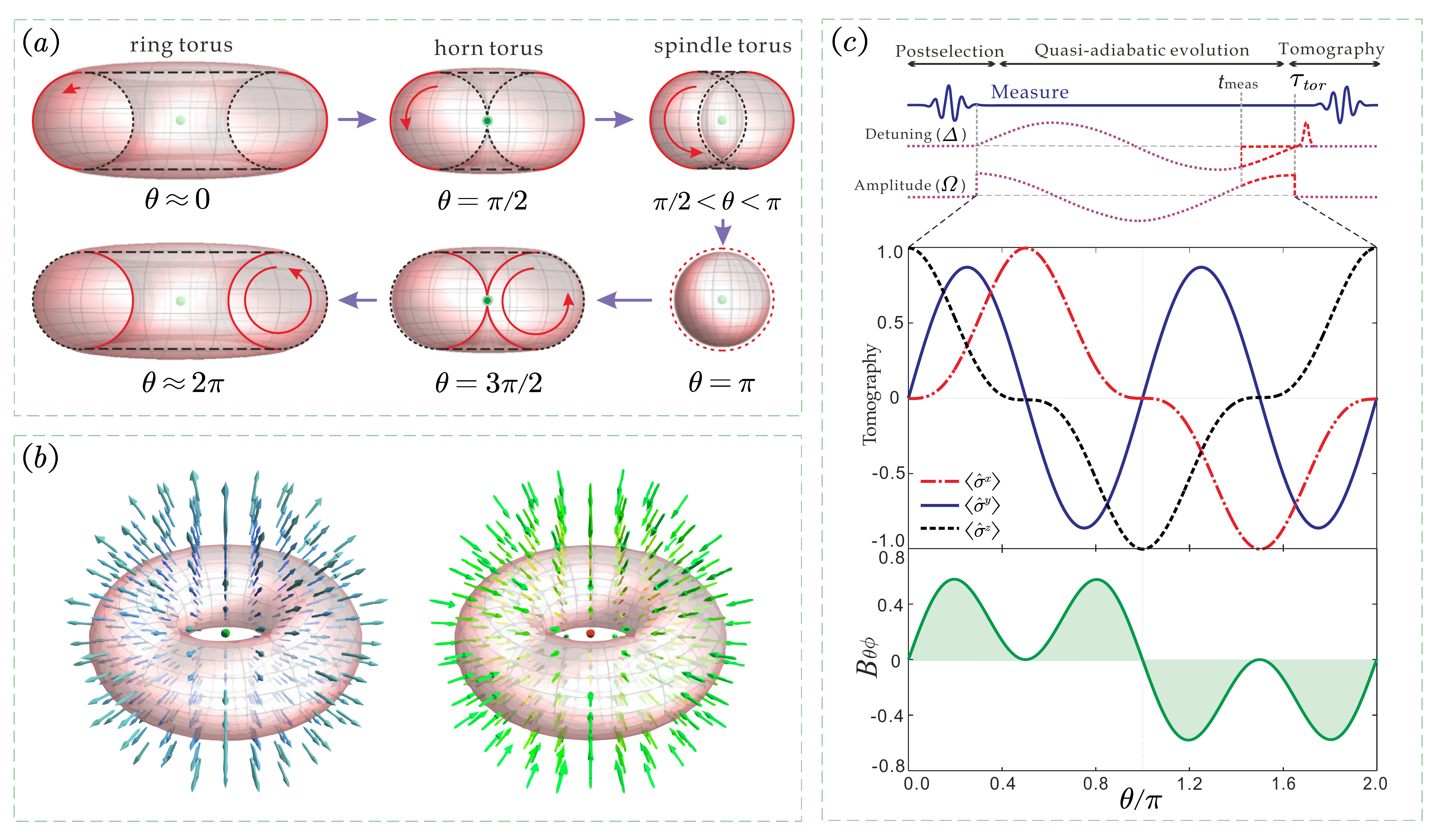}
    \caption{$($Color online$)$
    $($a$)$ As the distance to the origin decreases, a ring torus becomes a horn torus, then a spindle torus, and then degenerates into a sphere, and finally recurs to a ring torus. $($b$)$ Curves on the torus have different Gaussian curvatures, it could be greater than 0, equal to 0 and less than 0. But the Berry curvature is different. It is affected by the types of the quantum state manifolds: the ground state manifold for the sources $\rho_N$ $($the green ball$)$ and the excited state manifold for the sinks $\rho_S$ $($the red ball$)$. {$($c$)$} Measuring Berry curvature in the two-level system. The upper part describes the experimental pulse sequence. Following an initial measurement to project into the ground state, the Rabi frequency and the detuning are adjusted along a torus protocol, after which quantum state tomography is performed. The middle part depicts the process of such a tomography with parameters $\tau_{\textit{tor}}=1~\mu$s, $\Delta_1/2\pi=\Omega_1/2\pi=1.735~$MHz, and $\Delta_2=0$. The lower part represents the distribution of the Berry curvature during a complete period.}
    \label{fig:figure2}
  \end{figure*}

  Interestingly, if the locations of the Rabi frequency $\Omega$ and the detuning $\Delta$ are exchanged in the Hamiltonian matrix, the original manifold will transform into the torus parameter space. The Hamiltonian of the system is then given by
  \begin{eqnarray}\label{Hamiltonian Torus1}
  \hat{\textit{H}}_{\textrm{tor}}= \frac{1}{2}\left(
                   \begin{array}{cc}
                     \Omega & \Delta{\textit{e}}^{-\textit{i}\phi} \\
                     \Delta{\textit{e}}^{\textit{i}\phi} & -\Omega\\
                   \end{array}
                 \right).
  \end{eqnarray}

  The corresponding parametric equations of this Hamiltonian can be written as\footnote{While this paper was in preparation, we became aware of the interesting work published considering the similar Hamiltonian manifold transformation in Ref.~\cite{Mera-2019}.}
  \begin{eqnarray}\label{para_torus}
  \{\hat{\sigma}^x\}_{\textrm{tor}}=(\Delta_1\cos\theta+\Delta_2)\cos\phi,~~~~
  \{\hat{\sigma}^y\}_{\textrm{tor}}=(\Delta_1\cos\theta+\Delta_2)\sin\phi,~~~~
  \{\hat{\sigma}^z\}_{\textrm{tor}}=\Omega_1\sin\theta,
  \end{eqnarray}
  where $\{\cdot\}_{\textrm{tor}}$ has the same meaning in torus case. If the parameter $\Delta_2$ $($the radius of the orange circle in Fig.~\ref{fig:figure1}$($b$)$$)$ is changed from $-2\Delta_1$ to $2\Delta_1$, the torus will first contract and then expand around the degenerate point $($gapped region$)$ in the parameter space. As the distance to the gapped region decreases, the ring torus $(0 < \theta < \pi/2)$ becomes a horn torus $(\theta = \pi/2)$, then successively a spindle torus $(\pi/2 < \theta < 3\pi/2)$ and a sphere $(\theta = \pi)$, and finally recurs to the ring torus $(3\pi/2 < \theta < 2\pi)$ again within one period, as shown in Fig.~\ref{fig:figure2}$($a$)$.

\begin{figure}[t]
    \centering
    \includegraphics[width=4.2in]{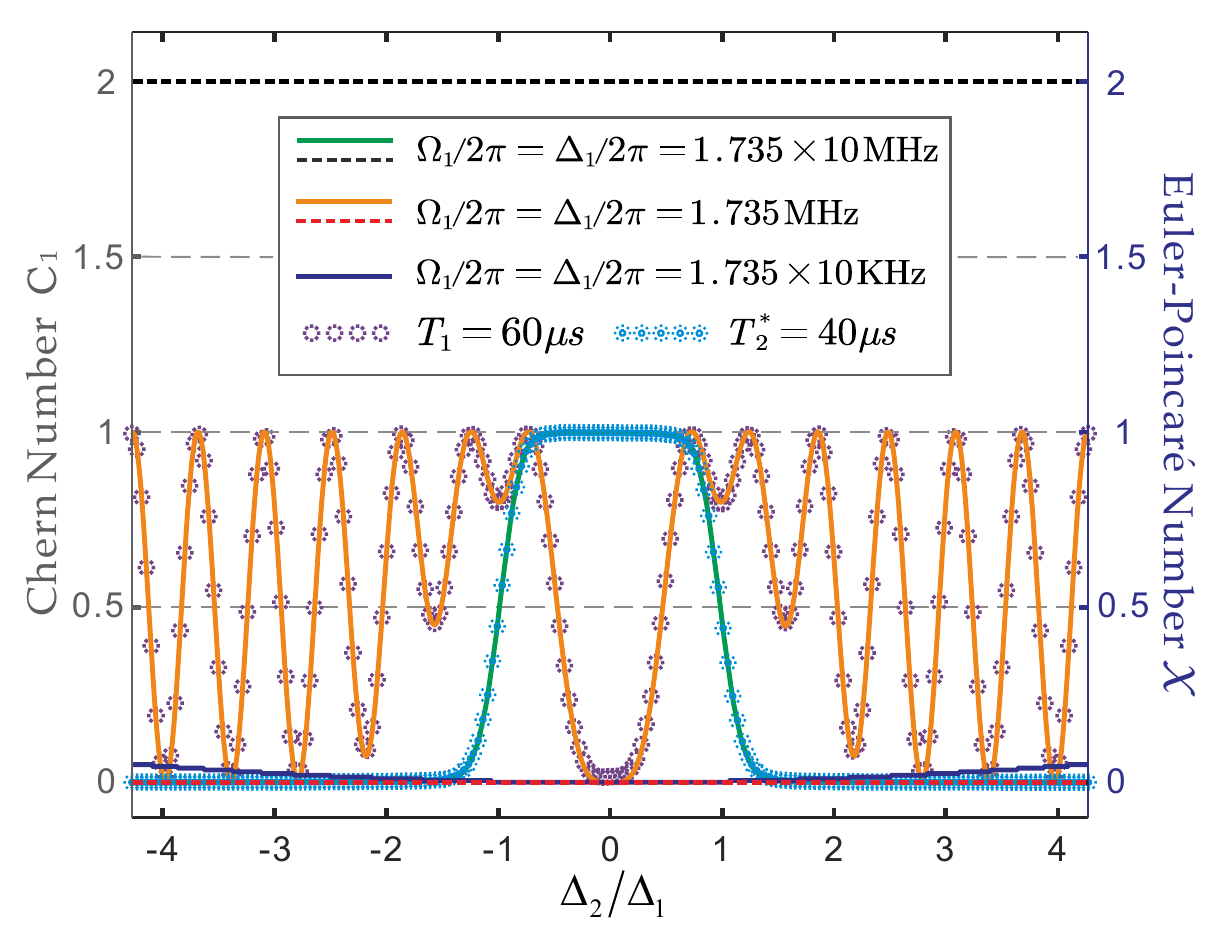}
    \caption{$($Color online$)$. Topological invariants. The first Chern number $\mathds{C}_1$ and the Euler-Poincar\'{e} characteristic number $\chi(\mathcal{M})=2(1-{g})$ for the sphere and the torus, respectively, where ${g}$ represents the genus of manifolds. The {green solid} line shows $\mathds{C}_1(\textrm{S}^2)$ and the {orange solid} line shows $\mathds{C}_1(\textrm{T}^2)$. The red dashed line shows $\chi(\textrm{T}^2)=0$ and the black dashed line shows $\chi(\textrm{S}^2)=2$. The numerical simulation using a Lindblad master equation is represented by the purple astragal and the cyan astragal. Here we notice that the {blue solid} line shows that the first Chern number of the torus with $\Delta_1/2\pi=\Omega_1/2\pi=1.735\times10~$kHz approximates to 0 and it will be explained in the next section.}
    \label{fig:figure3}
  \end{figure}

  We now analyze the distinctions between the Gaussian curvature and the Berry curvature in the Hamiltonian dynamics. Fig.~\ref{fig:figure2}$($b$)$ depicts the distribution of the Berry curvature around the torus. In the sight of differential geometry, according to Eqs.~$($\ref{para_sphere}$)$ and ~$($\ref{para_torus}$)$, the Gaussian curvatures of the sphere $\textit{K}_{\textrm{sph}}$ and the torus $\textit{K}_{\textrm{tor}}$ are separately given by~\cite{Struik-1961}
  \begin{eqnarray}\label{Gaussian1}
    \textit{K}_{\textrm{sph}}=\frac{1}{{\Omega_{1}^{2}}}
  \end{eqnarray}
  and
  \begin{eqnarray}\label{Gaussian2}
    \textit{K}_{\textrm{tor}}=\frac{\cos\theta}{\Delta_1(\Delta_2+\Delta_1\cos\theta)},
  \end{eqnarray}
  with the corresponding Gaussian area elements of the sphere and the torus being separately written as
   \begin{eqnarray}\label{elements of area1}
   d\sigma_{\textrm{sph}}=\Omega_1^2\sin\theta d\theta \wedge d\phi
   \end{eqnarray}
   and
   \begin{eqnarray}\label{elements of area2}
   d\sigma_{\textrm{tor}}={\Delta_1(\Delta_2+\Delta_1\cos\theta)}d\theta \wedge d\phi,
   \end{eqnarray}
  where $\Delta_1=\Omega_1$, and $\Delta_2$ varying from -4.27$\Delta_1$ to 4.27$\Delta_1$, which are depicted in Fig.~\ref{fig:figure1}$($b$)$. We thus then get the Euler-Poincar\'{e} characteristic number $\chi(\mathcal{M})=\frac{1}{2\pi}\int\textit{K}d\sigma$, as shown in Fig.~\ref{fig:figure3}. As we mentioned above, the degenerate points emerging from the Berry curvature $\textit{B}_{\mu\nu}$ act as sources and sinks of $\mathds{C}_1=\pm1$ that are analogous to Abelian monopoles in parameter space. In order to probe much deeper into the properties of the Berry curvature, we turn to the degenerate points at $\Delta=\Omega=0$ of the spherical coordinate around which these two manifolds can be manipulated. From the second term of Eq.~$($\ref{general force}$)$, the Berry curvatures of the sphere and the torus parameter space are given by
  \begin{eqnarray}\label{Berry curvature modify1}
  \textit{B}_{{\theta\phi}}^{\textrm{sph}}=\frac{\Omega_1\sin\theta}{2\theta_t^{\textrm{sph}}}
  \langle\hat{\sigma}^{y}\rangle
  \end{eqnarray}
  and
  \begin{eqnarray}\label{Berry curvature modify2}
  \textit{B}_{{\theta\phi}}^{\textrm{tor}}=\frac{\Delta_1\cos\theta+\Delta_2}{2\theta_t^{\textrm{tor}}}
  \langle\hat{\sigma}^{y}\rangle,
  \end{eqnarray}
  respectively, where $\theta_t^{\textrm{sph}}=\pi/{\tau_{\textrm{sph}}}$ is the quench velocity and ${\tau_{\textrm{sph}}}=1~\mu$s is the ramping time, of quantum state tomography in the sphere case~\cite{Schroer-2014}. For the torus case, we set $\theta_t^{\textrm{tor}}=2\pi/{\tau_{\textrm{tor}}}$ and $\tau_{\textrm{tor}}=1~\mu$s. The measurement period corresponds to $2\pi$ (before the ramping time ${\tau_{\textrm{tor}}}$) in the case of the torus manifold other than $\pi$ in the case of the spherical manifold, as shown in Fig.~\ref{fig:figure2}$($c$)$. Here, we set the torus of size $\Delta_1/2\pi=\Omega_1/2\pi=1.735$~MHz and $\Delta_2=0$.  As it is shown in Eq.~$($\ref{para_torus}$)$, the torus turns into a sphere with the radius $\Delta_1(\Omega_1)$.

   The Berry curvature can be extracted from the linear response of the qubit system to nonadiabatic manipulations $($deviations from adiabaticity$)$ of its Hamiltonian which leads to the first-order reaction force. As the Hamiltonians with parameters $\Delta$ and $\Omega$ are cylindrically symmetric about the $z$ axis, the corresponding Chern number $\mathds{C}_1(\mathcal{M})=(2\pi)^{-1}\int_{0}^{2\pi}d\phi\int_{0}^{\theta^{\prime}}d\theta\textit{B}_{\theta\phi}$ can be separately represented as
   \begin{eqnarray}\label{Chern of s}
   \mathds{C}_1(\textrm{S}^2)=\int_{0}^{\pi}\textit{B}_{\theta\phi}^{\textrm{sph}}d\theta
   \end{eqnarray}
   and
   \begin{eqnarray}\label{Chern of t}
   \mathds{C}_1(\textrm{T}^2)=\int_{0}^{2\pi}\textit{B}_{\theta\phi}^{\textrm{tor}}d\theta,
   \end{eqnarray}
   respectively, where $\theta^{\prime}=\pi$ for the sphere case and $\theta^{\prime}=2\pi$ for the torus case, similarly as we mentioned before.

   As shown in Fig.~\ref{fig:figure3}, we notice that the first Chern number of both manifolds are not constant. In the case of the sphere, the absolute value of the first Chern number $|\mathds{C}_1|=1$, if and only if the degenerate points are wrapped by the Hamiltonian manifold~\cite{Zhang1-2017}. However, the first Chern number of the torus is oscillating between 0 and 1, and the oscillation frequency is proportional to the magnitude of the parameters $($$\Delta$ and $\Omega$$)$. To some extent, the integrals in Eqs.~$($\ref{Chern of s}$)$ and $($\ref{Chern of t}$)$ are not $($quantized$)$ winding numbers when the manifolds are no longer enclosing the degenerate points $($gapped regions$)$. It is different from the result using the method calculating the conventional Berry curvature and Chern number~\cite{Hasan-2010}, as shown in Appendix~\ref{sec: concise comparison}. In the next section, we will notice that the first Chern number of a static torus actually approximates to zero, as shown in Fig.~\ref{fig:figure3}. Here, our choice of the magnitude of the parameters of the torus is ten times smaller than the sphere. The oscillation of the Chern number stems from the form of the Berry curvature in Eq.~$($\ref{Berry curvature modify2}$)$, with the condition of $\Delta_2$ varying from $-4.27\Delta_1$ to $4.27\Delta_1$. Under this condition, the first Chern number is not well defined.

   To study the robustness of manipulating the qubit dynamics, we need to put our proposal into an open quantum system. The whole dynamics of the system decaying to the reservoir can be modeled by the Lindblad master equation
  \begin{eqnarray}\label{Hamiltonian Torus2}
  \frac{d\hat{\rho}(t)}{dt}=i[\hat{\rho}(t),\hat{{\textit{H}}}_{\textrm{tor}}(t)]+\frac{\mathcal{D}[\hat{\sigma}_-]\hat{\rho}(t)}{T_1}+\frac{\mathcal{D}
  [\hat{\sigma}_{z}]\hat{\rho}(t)}{2T_{\phi}},
  \end{eqnarray}
  where we use Pauli operator $\hat{\sigma}_{z}$, annihilation operator $\hat{\sigma}_{-}=\hat{\sigma}_{x}-i\hat{\sigma}_{y}$, and the Lindblad super-operator is defined for any operator $\hat{\mathcal{L}}$ as $\mathcal{D}[\hat{\mathcal{L}}]\hat{\rho}=\hat{\mathcal{L}}\hat{\rho}\hat{\mathcal{L}}^{\dag}-(1/2)\hat{\mathcal{L}}^{\dag}\hat{\mathcal{L}}\hat{\rho}-(1/2)
  \hat{\rho}\hat{\mathcal{L}}^{\dag}\hat{\mathcal{L}}$. The qubit dephasing rate is $1/\textit{T}_{\phi}=1/\textit{T}_{2}^{*}-1/2\textit{T}_{1}$, where $\textit{T}_{\phi}$ represents the dephasing time, $\textit{T}_{1}$ is the relaxation time characterizing a qubit relaxing from the excited state $|e\rangle$ to the ground state $|g\rangle$, and $\textit{T}_{2}^{*}$ depicts the Ramsey dephasing time. The system's parameters we take for the simulation are $\textit{T}_1=60~{\mu}$s and $\textit{T}_{2}^{*}=40~{\mu}$s, which are about to be available with the improvement of
  the recently implemented superconducting devices~\cite{Song-2017,Ning-2019}.

\section{The connection between the Berry curvature and the quantum states evolution}
\label{sec: The connection between the Berry curvature and the quantum state evolution}

  Now, let us show the connection between the Berry curvature of the torus Hamiltonian parameter space and the fidelity of system's states. For simplicity, here we just consider the case of the source $\rho_N$ which sits at $\Delta=\Omega=0$ in the parameter space. The eigenstates of the Hamiltonian $\hat{\textit{H}}_{\textrm{tor}}$ are given by
  \begin{eqnarray}\label{Eigenstates1}
  |\psi_e\rangle=\frac{\Delta/2|e\rangle}{\sqrt{{{\Delta}^2}/{4}+(\textit{E}_e-{\Omega}/{2})^2}}
  +\frac{e^{i\phi}(\textit{E}_e-{\Omega}/{2})|g\rangle}{\sqrt{{{\Delta}^2}/{4}+(\textit{E}_e-{\Omega}/{2})^2}}
  \end{eqnarray}
  and
  \begin{eqnarray}\label{Eigenstates2}
  |\psi_g\rangle=\frac{\Delta/2|e\rangle}{\sqrt{{{\Delta}^2}/{4}+(\textit{E}_g-{\Omega}/{2})^2}}
  +\frac{e^{i\phi}(\textit{E}_g-{\Omega}/{2})|g\rangle}{\sqrt{{{\Delta}^2}/{4}+(\textit{E}_g-{\Omega}/{2})^2}},
  \end{eqnarray}
  respectively, with the eigenvalues $\textit{E}_{e(g)}=\pm{\sqrt{{\Delta}^2+{\Omega}^2}}/{2}$. The fidelity for the quantum state is defined as $f=\langle\psi_\alpha|\hat{\rho}(t_f)|\psi_\alpha\rangle$ $($$\alpha=e,g$$)$, with the target state $\hat{\rho}(t_f)=|g\rangle\langle g|$. The system's states can then be tracked by the change of the Berry curvature.

  \begin{figure}[t]
    \centering
    \includegraphics[width=7in]{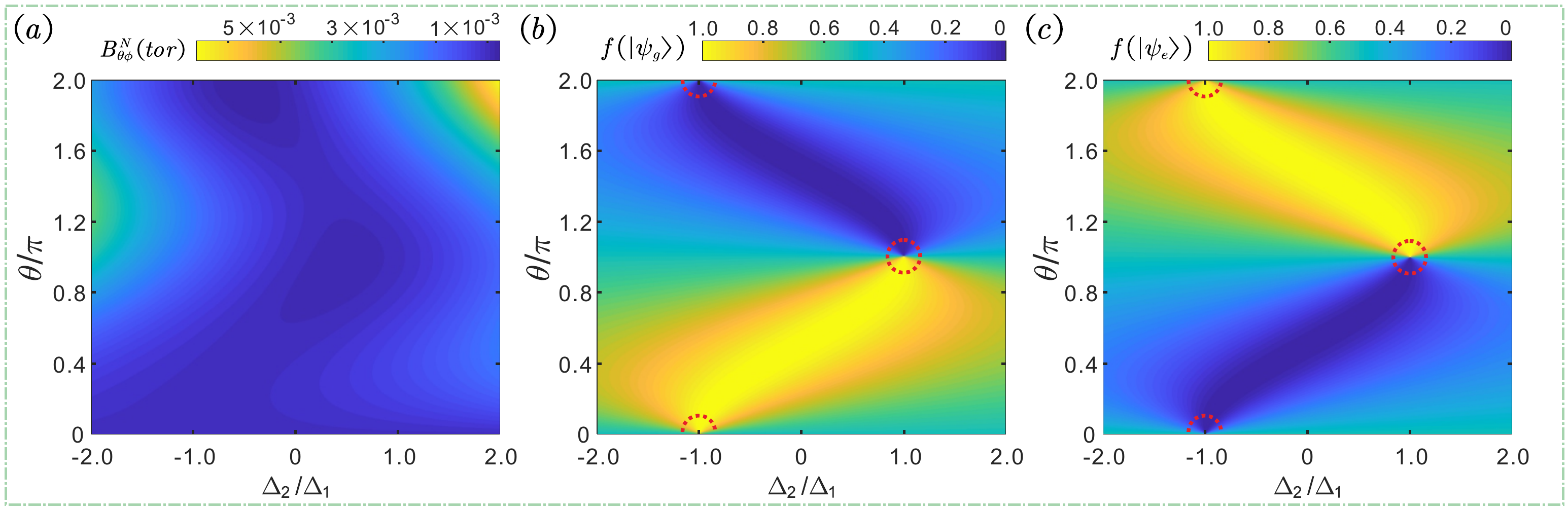}
    \caption{$($Color online$)$ $($a$)$ The Berry curvature measured as a function of $\Delta_2/\Delta_1$ for which we take the values of the parameters $\Delta_1/2\pi=\Omega_1/2\pi=1.735\times10~$kHz, $\Delta_2$ from $-2\Delta_1$ to $2\Delta_1$, and the ramping time $\tau_{\textrm{tor}}=1~\mu$s. The region of the Berry curvature is from $-2.375\times10^{-4}$ to $5.623\times10^{-3}$. $($b$)$ The fidelity: ${|\langle g|\psi_g\rangle|}^2$. $($c$)$ The fidelity: ${|\langle g|\psi_e\rangle|}^2$. The dashed red circles in $($b$)$ and $($c$)$ enclose the singularities which are located at $[\Delta_2/\Delta_1=-1, \theta/\pi=0], [\Delta_2/\Delta_1=-1, \theta/\pi=2]$, and $[\Delta_2/\Delta_1=1, \theta/\pi=1]$, respectively.}
    \label{fig:figure4}
  \end{figure}

  In order to manifest the physical meaning that stems from the Berry curvature, we first consider the adiabatic case. If we set the parameters $\Delta_1/2\pi=\Omega_1/2\pi=1.735\times10~$kHz, from Fig.~\ref{fig:figure2} $($b$)$, we can imagine that the strength of the ``magnetic fields'' originated from the singularities are too small to have a significant effect on the evolution of the system's states, as shown in Fig.~\ref{fig:figure4}. Under the circumstances, the corresponding first Chern number of the torus no longer oscillates and approximates to zero $($the first Chern number of the torus is approximately equal to the Euler-Poincar\'{e} characteristic number here$)$, as mentioned in the last section of Fig.~\ref{fig:figure3}. The system's states evolve adiabatically in their own eigenstate spaces. In Fig.~\ref{fig:figure4}$($b$)$ and ~\ref{fig:figure4}$($c$)$, we notice that each of them has three interesting points enclosed by the red dotted circles. Actually, these points correspond to the same singularity in the torus parameter space. As shown in Fig.~\ref{fig:figure2}$($a$)$, the spherical manifold emerges twice during the process of the deformation of the torus for $\Delta_2$ varying from $-2\Delta_1$ to $2\Delta_1$. The transition of the system's states occurs at $|\Delta_2/\Delta_1|=1$ during the process with $\Delta_2$ varying from $-\Delta_1$ to $\Delta_1$.

\begin{figure*}[t]
    \includegraphics[width=7in]{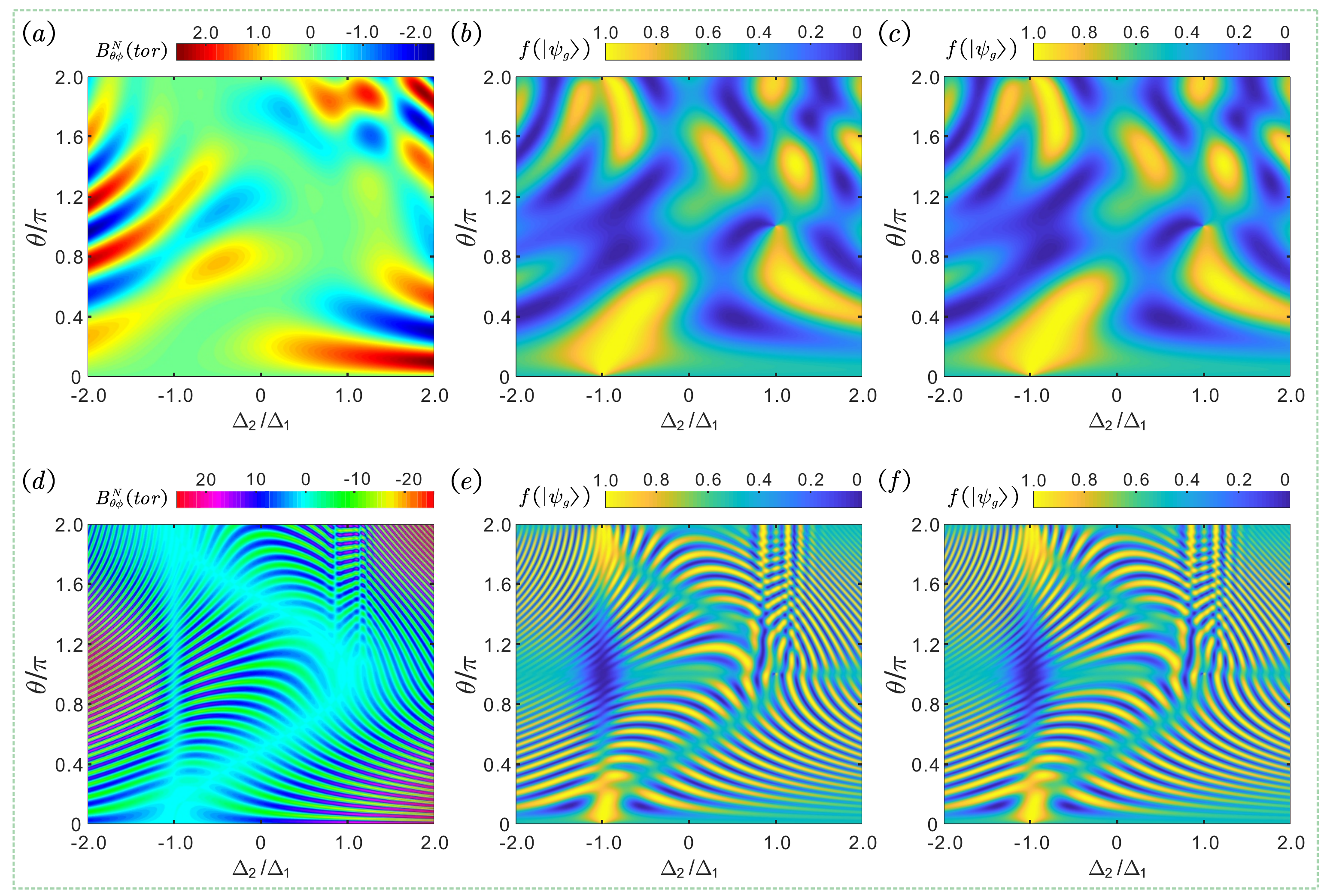}
    \caption{\label{fig3}$($Color online$)$. $($a$)$ The Berry curvature measured as a function of $\Delta_2/\Delta_1$ for which we take the values of the parameters $\Delta_1/2\pi=\Omega_1/2\pi=1.735~$MHz, $\Delta_2$ from $-2\Delta_1$ to $2\Delta_1$, and the ramping time $\tau_{\textit{tor}}=1~\mu$s. With the same setting, $($d$)$ is only different with $\Delta_1/2\pi=\Omega_1/2\pi=1.735\times10~$MHz. $($b$)$ and $($e$)$ depict the fidelity of $|\psi_g\rangle$ without qubit's decoherence. $($c$)$ and $($f$)$ represent the fidelity of $|\psi_g\rangle$ with $\textit{T}_1=60{\mu}$s and $\textit{T}_{2}^{*}=40{\mu}$s. The density matrix of the target state here is $\hat{\rho}(t_f)=|g\rangle\langle g|$.}
    \label{fig:figure5}
  \end{figure*}

  It becomes aware of the interesting connection between the Berry curvature and the evolution of the system's states by controlling the system's parameters. In Fig.~\ref{fig:figure5}, the Berry curvature ${\textit{B}}_{\theta\phi}$ and the fidelity of the target state $|g\rangle$ are plotted versus $\theta/\pi$ and $\Delta_2/\Delta_1$.
  If we set $\Delta_1/2\pi={\Omega_1/2\pi}=1.735~$MHz, we can realize that one of the eigenstates $|\psi_g\rangle$ is slightly affected by the corresponding Berry curvature. Both of the eigenstates $(|\psi_g\rangle$ and $|\psi_e\rangle)$ no longer evolve in the adiabatic way. The deviations from the adiabaticity become large enough to affect the evolution of the system's states, as shown in Fig.~\ref{fig:figure5}. We continue to tune up the parameters to $\Delta_1/2\pi={\Omega_1/2\pi}=1.735\times10~$MHz, where the Berry curvature becomes ten times larger than before.

  From Ref.~\cite{Zhang1-2017}, we know that the magnitude of the Berry curvature in parameter space corresponds to the strength of the magnetic field of the monopoles $($singularities$)$. When the degenerate points pass through the parameter space manifolds, the quantum states influenced by the Berry curvature will cause ripples in the Hilbert space, this phenomenon is more obvious in Fig.~\ref{fig:figure5}$($e$)$ than in Fig.~\ref{fig:figure5}$($b$)$ for $\Delta_2/\Delta_1=1$. In Fig.~\ref{fig:figure5}$($d$)$, we notice that there is a significantly closed triangular region emerging from $\Delta_2/\Delta_1=-1$ to $\Delta_2/\Delta_1=1$. This reflects the process of the measurement of the Berry curvature. Actually, the triangular region also exists in Fig.~\ref{fig:figure4}$($a$)$ and Fig.~\ref{fig:figure5}$($a$)$. However, the magnitude of the Berry curvature is too small to show the region. In Fig.~\ref{fig:figure5}$($c$)$ and \ref{fig:figure5}$($f$)$, the numerical simulation using the Lindblad master equation of Eq.~$($\ref{Hamiltonian Torus2}$)$ models the imperfect dynamics with $\textit{T}_1=60~{\mu}$s and $\textit{T}_{2}^{*}=40~{\mu}$s, showing that the fidelity of the system's states almost stay unchanged under the considered noise effects.

\section{Concluding remarks}
\label{sec: Concluding remarks}

  In conclusion, we study the Berry curvature characterized by Hamiltonian parameter space and its connection to the evolution of the system's states with a microwave-addressed two-level system. We firstly show two different topological Hamiltonian parameter spaces, a sphere and a torus. The dynamics revealed is, to an extent, to simulate that, strength of the magnetic
  field described by the Berry curvature influences the evolution of the system's states during the whole process of the deformation of the manifold.

Generally, the first Chern number is always an \textbf{integer number} if the base space is a \textbf{compact oriented two-dimensional smooth manifold}.
  However, the first Chern number of the variational torus manifold is no more discrete numbers under the circumstance in the case considered here. We need to emphasize that the non-integer first Chern number mentioned in the dynamically non-adiabatic response method is due to an artifact of approximating curvature.
  Discrete numbers still can be obtained by considering an irreducible representation of SU$(2)$ with spin-1 case~\cite{Mera-2019}. However, this is not the key point here. Back to our case, in the adiabatic case, the first Chern number of the torus is approximately equivalent to the Euler-Poincar\'{e} number. With the increase of the magnitude of the Berry curvature, the balance between them is gradually broken.

  For different kinds of manifolds with different geometrical and topological structures in parameter space, different distributions of the Berry curvature will affect variously the evolution of quantum states. We find that Berry curvature plays a crucial role in connecting the Hamiltonian parameter space and the Hilbert space. We can built up some other kinds of Hamiltonian manifolds in parameter space. This helps us to gain insight into the manipulation of the qubit dynamics from the geometric point of view.
\begin{acknowledgments}

 This work was supported in part by the National Natural Science Foundation of China under Grants No.11875108, No.11405031, and the Natural Science Foundation of Fujian Province under Grants 2018J01412.

\end{acknowledgments}

\begin{appendix}
\section{A concise comparison between two approaches for calculating the Berry curvature}
\label{sec: concise comparison}

In general case, the first Chern number of the ground state manifold is encoded in the gauge-invariant curvature 2-form which can be written as~\cite{Hasan-2010}
  \begin{eqnarray}\label{gauge-invariant}
  {\mathcal{B}=\textit{B}_{{\theta\phi}}d\theta\wedge{d}\phi,}
  \end{eqnarray}
where $\textit{B}_{{\theta\phi}}$ is the conventional Berry curvature. For our two-level Hamiltonian, the conventional Berry curvature of the ground state in the gapped region can be expressed as
  \begin{eqnarray}\label{Berry 2-form}
  \textit{B}_{{\theta\phi}}=\frac{1}{2}\hat{x}\cdot(\frac{\partial\hat{x}}{\partial\theta}\times\frac{\partial\hat{x}}{\partial\phi})
  \end{eqnarray}
via the normalized vector component $\hat{x}={\pmb{x}}/\|\pmb{x}\|$, which in the sphere and the torus cases are respectively
 \begin{eqnarray}\label{ps}
  \pmb{x}_{\textrm{sph}}&=&({x}^{1},{x}^{2},{x}^{3})_{\textrm{sph}}
  =\{\hat{\sigma}^{\mu}\}_{\textrm{sph}}
  =\{\hat{\sigma}^{x},\hat{\sigma}^{y},\hat{\sigma}^{z}\}_{\textrm{sph}}=(\Omega\cos\phi,~\Omega\sin\phi,~\Delta)
 \end{eqnarray}
 and
  \begin{eqnarray}\label{pt}
  \pmb{x}_{\textrm{tor}}&=&({x}^{1},{x}^{2},{x}^{3})_{\textrm{tor}}
  =\{\hat{\sigma}^{\mu}\}_{\textrm{tor}}
  =\{\hat{\sigma}^{x},\hat{\sigma}^{y},\hat{\sigma}^{z}\}_{\textrm{tor}}=(\Delta\cos\phi,~\Delta\sin\phi,~\Omega).
 \end{eqnarray}
 We can then calculate the conventional Berry curvatures of the two families via Eq.~$($\ref{Berry 2-form}$)$ to be
  \begin{eqnarray}\label{Berry 2-form sphere}
  \textit{B}_{{\theta\phi}}^{\rm{S}^2}=\frac{\Omega_1^2\sin\theta(\Delta_2\cos\theta+\Delta_1)}{2\big[(\Delta_1\cos\theta+\Delta_2)^2+\Omega_1^2\sin^2\theta\big]^{3/2}}
  \end{eqnarray}
  and
  \begin{eqnarray}\label{Berry 2-form torus}
  \textit{B}_{{\theta\phi}}^{\rm{T}^2}=-\frac{\Omega_1(2(\Delta_1^2+\Delta_2^2)\cos\theta+\Delta_2 \Delta_1(\cos2\theta+3))}{4\big[(\Delta_1\cos\theta+\Delta_2)^2+\Omega_1^2\sin^2\theta\big]^{3/2}}.
  \end{eqnarray}
  The corresponding first Chern number of them can be calculated as
  \begin{eqnarray}\label{Chern of sp}
   \mathds{C}_1(\textrm{S}^2)=\frac{1}{2\pi}\int_{0}^{\pi}d\theta\int_{0}^{2\pi}{d}\phi~\textit{B}_{\theta\phi}^{\rm{S}^2}
   =\frac{1}{2}\left(\frac{\Delta_1-\Delta_2}{\sqrt{\left(\Delta_1-\Delta_2\right){}^2}}+\frac{\Delta_1+\Delta _2}{\sqrt{\left(\Delta_1+\Delta_2\right){}^2}}\right)
  \end{eqnarray}
   and
  \begin{eqnarray}\label{Chern of to}
   \mathds{C}_1(\textrm{T}^2)=\frac{1}{2\pi}\int_{0}^{2\pi}d\theta\int_{0}^{2\pi}{d}\phi~\textit{B}_{\theta\phi}^{\rm{T}^2}=0.
  \end{eqnarray}
  From Eq.~$($\ref{Chern of sp}$)$, we notice that the first Chern number is not well defined for $\Delta_1=\pm\Delta_2$.

  We now numerically calculate the Berry curvature using the dynamical method as follows ~\cite{Gritsev-2012}:
  \begin{eqnarray}\label{Berry curvature m1}
  \textit{B}_{{\theta\phi}}^{\textrm{sph}}&=&\frac{\langle\psi(t)|\partial_{\phi}\hat{\textit{H}}_{\textrm{sph}}|\psi(t)\rangle}{\theta_t^{\textrm{sph}}}
  =\frac{1}{2\theta_t^{\textrm{sph}}}
  \bigg\langle\psi(t)\bigg|\partial_{\phi}\left(
                   \begin{array}{cc}
                     \Delta & \Omega{\textit{e}}^{-\textit{i}\phi} \\ 
                     \Omega{\textit{e}}^{\textit{i}\phi} & -\Delta\\
                   \end{array}
                 \right)\bigg|\psi(t)\bigg\rangle
                 =\frac{\Omega_1\sin\theta}{2\theta_t^{\textrm{sph}}}
  \bigg\langle\psi(t)\bigg|\left(
                   \begin{array}{cc}
                     0 & -i \\
                     i & 0\\
                   \end{array}
                 \right)\bigg|\psi(t)\bigg\rangle\\ \cr\cr
                 &=&\frac{\Omega_1\sin\theta}{2\theta_t^{\textrm{sph}}}
  \langle\hat{\sigma}^y\rangle
  =\frac{\Omega_1\sin\theta}{2\theta_t^{\textrm{sph}}}
  \textrm{Tr}(\hat{\rho}\hat{\sigma}^y)
  \end{eqnarray}
  and
  \begin{eqnarray}\label{Berry curvature m2}
  \textit{B}_{{\theta\phi}}^{\textrm{tor}}&=&\frac{\langle\psi(t)|\partial_{\phi}\hat{\textit{H}}_{\textrm{tor}}|\psi(t)\rangle}{\theta_t^{\textrm{tor}}}
  =\frac{1}{2\theta_t^{\textrm{tor}}}
  \bigg\langle\psi(t)\bigg|\partial_{\phi}\left(
                   \begin{array}{cc}
                     \Omega & \Delta{\textit{e}}^{-\textit{i}\phi} \\
                     \Delta{\textit{e}}^{\textit{i}\phi} & -\Omega\\
                   \end{array}
                 \right)\bigg|\psi(t)\bigg\rangle
                 =\frac{\Delta_1\cos\theta+\Delta_2}{2\theta_t^{\textrm{tor}}}
  \bigg\langle\psi(t)\bigg|\left(
                   \begin{array}{cc}
                     0 & -i \\
                     i & 0\\
                   \end{array}
                 \right)\bigg|\psi(t)\bigg\rangle~~~~\\  \cr\cr
                 &=&\frac{\Delta_1\cos\theta+\Delta_2}{2\theta_t^{\textrm{tor}}}\langle\hat{\sigma}^y\rangle
                 =\frac{\Delta_1\cos\theta+\Delta_2}{2\theta_t^{\textrm{tor}}}
  \textrm{Tr}(\hat{\rho}\hat{\sigma}^y),
  \end{eqnarray}
 where $\hat{\rho}$ is the density matrix of the lower dressed state $|\psi(t)\rangle=|\psi_g\rangle$ in Eq.~$($\ref{Eigenstates2}$)$. Eqs.~$($\ref{general force}$)$, $($\ref{Berry curvature m1}$)$ and $($\ref{Berry curvature m2}$)$ show the approach to measure Berry curvature is a leading order approximation, good to the first order in the ramping rate $\theta_{t}$ $($sphere case$)$. However, we notice that the first Chern number in the torus case is no longer a constant number
 outside the gapped region. The various oscillations in Fig.~\ref{fig:figure3} and Fig.~\ref{fig:figure5} noted in the text could not happen in the case of
 the conventional Berry curvature. From Ref.~\cite{Gritsev-2012}, we know that the method is not that perfect when the ramp velocity is sufficient large $($large deviation from the linear response regime$)$. Therefore, the dynamically non-adiabatic response method we adopt here actually is not equivalent to the conventional one under certain
 circumstances.
 \end{appendix}


\begin{thebibliography}{999}

   \bibitem{Berry-1984} M. V. Berry, Quantal phase factors accompanying adiabatic changes, \textrm{Proc. R. Soc. A}, \textbf{392}, 45 (1984).

   \bibitem{Gritsev-2012} V. Gritsev and A. Polkovnikov, Dynamical quantum Hall effect in the parameter space, \textrm{Proc. Natl. Acad. Sci. U.S.A.}, \textbf{109}, 6457 (2012).

   \bibitem{Hasan-2010} M. Z. Hasan and C. L. Kane, Topological insulators, \textrm{Rev. Mod. Phys.}, \textbf{82}, 3045 (2010).

   \bibitem{Sugawa-2018} S. Sugawa, F. Salces-Carcoba, P. R. Perry, Y. C. Yue, I. B. Spielman, Second Chern number of a quantum-simulated non-Abelian Yang monopole, \textit{Science}, \textbf{360}, 6396 (2018).

   \bibitem{Flaschner-2016} N. Fl$\ddot{\mathrm{a}}$schner, B. S. Remet al., Experimental reconstruction of the Berry curvature in a Floquet Bloch band, \textrm{Science}, \textbf{352}, 6289 (2016).

   \bibitem{Provost-1980} J. Provost and G. Vallee, Riemannian structure on manifolds of quantum states, \textrm{Commun. Math. Phys.}, \textbf{76}, 289 (1980).

   \bibitem{Simon-1983} B. Simon, Holonomy, the Quantum Adiabatic Theorem, and Berry's Phase, \textrm{Phys. Rev. Lett.}, \textbf{51}, 2167 (1983).

   \bibitem{Zhang1-2017} Z. L. Zhang, M. F. Chen, H. Z. Wu and Z. B. Yang, Quantum simulation of Abelian Wu-Yang monopoles in spin-1/2 systems, \textrm{Laser Phys. Lett.}, \textbf{14}, 045204 (2017).

   \bibitem{Zhang2-2017} Z. L. Zhang, M. F. Chen, H. Z. Wu and Z. B. Yang, Quantum simulation of gravitational-like waves in minisuperspace with an artificial qubit, \textrm{Phys. Rev. D}, \textbf{95}, 046010 (2017).

   \bibitem{Schroer-2014} M. D. Schroer, M. H. Kolodrubetzet et al., Measuring a topological transition in an artificial spin-1/2 system, \textrm{Phys. Rev. Lett.}, \textbf{113}, 050402 (2014).

   \bibitem{Berry-1993} M. V. Berry, and J. M. Robbins, Chaotic classical and half-classical adiabatic reactions: geometric magnetism and deterministic friction, \textrm{Proc. R. Soc.}, \textbf{442}, 659 (1993).

   \bibitem{Roushan-2014} P. Roushan, C. Neill et al., Observation of topological transitions in interacting quantum circuits, \textrm{Nature}, \textbf{515}, 241 (2014).

   \bibitem{Struik-1961} D. J. Struik, \textit{Lectures on Classical Differential Geometry}, (Addison-Wesley Pub. Co., 1961).

   \bibitem{Mera-2019} B. Mera, K. Sacha, and Y. Omar, Topologically Protected Quantization of Work, \textrm{Phys. Rev. Lett.}, \textbf{123}, 020601 (2019).

   \bibitem{Song-2017} C. Song, S. B. Zheng, et al., Continuous-variable geometric phase and its manipulation for quantum computation in a superconducting circuit, \textrm{Nat. Commun.}, \textbf{8}, 1061 (2017).

   \bibitem{Ning-2019} W. Ning, X. J. Huang et al., Deterministic Entanglement Swapping in a Superconducting Circuit, \textrm{Phys. Rev. Lett.}, \textbf{123}, 060502 (2019).

\end{thebibliography}
\end{document}